\documentclass{aa}  

\usepackage[hyperindex, breaklinks, colorlinks, linkcolor=blue, citecolor=blue]{hyperref}
\usepackage{graphicx}
\usepackage{upgreek}
\usepackage{svg}
\usepackage{orcidlink}
\usepackage{calrsfs}
\DeclareMathAlphabet{\pazocal}{OMS}{zplm}{m}{n}

\newcommand{\citeyearless}[1]{\citeauthor{#1} \citeyear{#1}}
\def\mean#1{\left< #1 \right>}
\begin{document}

   \title{A fast and robust recipe for modeling non-ideal MHD effects in star-formation simulations}

   \author{E. Agianoglou\inst{1, 2}\orcidlink{0009-0007-7334-3193},
          A. Tritsis\inst{3}\orcidlink{0000-0003-4987-7754}
          \fnmsep\thanks{aris.tritsis@epfl.ch}
          \and
          K. Tassis\inst{1, 2}\orcidlink{0000-0002-8831-2038}}

   \institute{Department of Physics, and Institute for Theoretical and Computational Physics, University of Crete, Voutes University campus, GR-70013 Heraklion, Greece 
        \and
            Institute of Astrophysics, Foundation for Research and Technology-Hellas, N. Plastira 100, Vassilika Vouton, GR-71110 Heraklion, Greece
         \and
            Institute of Physics, Laboratory of Astrophysics, Ecole Polytechnique F\'ed\'erale de Lausanne (EPFL), \\ Observatoire de Sauverny, 1290, Versoix, Switzerland}

   \date{Received date; accepted date}
   \titlerunning{Fast modeling of non-ideal MHD effects}
   \authorrunning{Agianoglou et al.}
 
  \abstract
   {Non-ideal magnetohydrodynamic (MHD) effects are thought to be a crucial component of the star-formation process. Numerically, several complications render the study of non-ideal MHD effects in 3-dimensional (3D) simulations extremely challenging and hinder our efforts of exploring a large parameter space.}
   {Here, we aim to overcome such challenges by proposing a novel, physically-motivated empirical approximation to model non-ideal MHD effects.}
   {We perform a number of 2D axisymmetric 3-fluid non-ideal MHD simulations of collapsing prestellar cores and clouds with non-equilibrium chemistry and leverage upon previously-published results from similar simulations with different physical conditions. We utilize these simulations to develop a multivariate interpolating function to predict the ionization fraction in each region of the cloud depending on the local physical conditions. We subsequently use analytically-derived, simplified expressions to calculate the resistivities of the cloud in each grid cell. Therefore, in our new approach the resistivities are calculated without the use of a chemical network. We benchmark our method against additional 2D axisymmetric non-ideal MHD simulations with random initial conditions and a 3D non-ideal MHD simulation with non-equilibrium chemistry.}
    {We find excellent quantitative and qualitative agreement between our approach and the ``full'' non-ideal MHD simulations both in terms of the spatial structure of the simulated clouds and regarding their time evolution. At the same time, we achieve a factor of $\sim10^2-10^3$ increase in computational speed. Given that we ignore the contribution of grains to the resistivities our approximation is valid up to number densities of $\sim10^6 ~\rm{cm^{-3}}$ and is therefore suitable for pc-scale simulations of molecular clouds and/or simulations of stratified boxes. The tabulated data required for integrating our method in hydrodynamical codes, along with a \textsc{fortran} implementation of the interpolating function are publicly available \href{https://github.com/manosagian/Non-Ideal-MHD-Approximate-Code}{here}.}
   {}


   \maketitle


\section{Introduction}\label{intro}

Non-ideal magnetohydrodynamic (MHD) effects, and particularly ambipolar diffusion, play a crucial role
in star formation (e.g., \citeyearless{1999ASIC..540..305M}). They allow the collapse of cores with
a magnetic field initially strong enough to prevent collapse by redistributing the magnetic flux in the cloud
(e.g., \citeyearless{2009NewA...14..483B}; \citeyearless{2010MNRAS.408..322K}; \citeyearless{2012ApJ...754....6T}; \citeyearless{2022MNRAS.510.4420T}). Such a
redistribution of magnetic flux is necessary to explain why stars do not inherit all of the magnetic flux of their parent clouds, thereby providing a solution to the ``magnetic-flux problem'' in star formation (e.g., \citeyearless{2015MNRAS.452..278T}). Finally, a reduced coupling between the magnetic field and the gas is required to mitigate the effect of magnetic breaking, allowing rotationally-supported disks to form (see e.g., \citeyearless{2020MNRAS.495.3795W} and references therein).

Numerically, there are two major complications when modeling non-ideal MHD effects.
The first challenge is that, in explicit codes, the maximum allowed time step for stability, determined by the Courant–Friedrichs–Lewy (CFL) condition \citep{1928MatAn.100...32C}, needs to be reduced by many orders of magnitude in comparison to the ideal MHD regime, in order to account for the diffusion processes introduced (\citeyearless{1995ApJ...442..726M}; \citeyearless{2001ApJ...550..314D}). This can lead to prohibitively small timesteps and thus very computationally-expensive simulations. The second major issue stems from the need to include a chemical network in order to calculate the abundances of ions from which the resistivities can then be self-consistently calculated. This often involves solving hundreds to thousands of chemical reactions in every cell in the simulation box and for every time step. These two issues become especially problematic for high-resolution, 3-dimensional (3D) simulations, therefore hindering our efforts of studying such effects in 3D and exploring fully the parameter-space.

Several attempts have been made over the past couple of decades to mitigate these issues. \cite{2006ApJ...653.1280L} proposed the so-called ``heavy-ion approximation'', in which the time step is increased while maintaining the ambipolar diffusion timescale. However, this approach is not physically motivated as the mass of the ions is artificially increased, and its applicability to turbulent clouds is problematic. In other studies (e.g., \citeyearless{2024ApJ...961..100A}, \citeyearless{2022arXiv220913765T}), the calculation of the resistivities is simplified by using simple power-law scaling with the density for either the ion density or the resistivities. However, it remains ambiguous to what extent the scaling of the resistivities with the gas density can be accurately approximated by a single power law throughout the evolution of a cloud.

Here, we aim to solve the second major issue when performing non-ideal MHD simulations; that is the need to include a chemical network. While we also use a power-law for approximating the ionization fraction in the cloud as a function of the $\rm{H_2}$ density, in our implementation the power law evolves alongside the collapsing cloud. This approximation and its evolution depend on the physical conditions of any given model. Therefore, an interpolation function over the physical parameters inside the cloud (temperature, cosmic-ray ionization rate, visual extinction etc.) is used to generalize the method for a large parameter space. With this new method, the resistivities of a cloud can be accurately calculated for various physical conditions without the need to calculate any chemical reactions.

We note here that the contribution of grains in the resistivities is not taken into account in this new approach. Consequently, our approximations are valid up to a density of $\sim10^6 ~\rm{cm^{-3}}$ and our method is specifically targeted at pc-scale star formation simulations that focus on the fragmentation and core formation phase (\citeyearless{2014MNRAS.444.2396C}, \citeyearless{2018PASJ...70S..53I}, \citeyearless{2019MNRAS.488.1407K}, \citeyearless{2023ApJ...954...93A}). Such simulations typically incorporate sink particles at densities $10^7 ~\rm{cm^{-3}}$, effectively bypassing the need for a detailed treatment of non-ideal MHD effects beyond these densities. Therefore, our method provides an efficient and practical way to include non-ideal MHD effects for the majority of the time evolution of such numerical studies.

This paper is organized as follows: In Sect.~\ref{eqs} we derive approximate expressions for the perpendicular, parallel and Hall resistivities. In Sect.~\ref{numsetup} we describe the setup of 2D and 3D non-ideal MHD (chemo-)dynamical simulations of collapsing molecular clouds and cores used to develop our method and benchmark our results. In Sect.~\ref{results} we present our findings and in Sect.~\ref{sum} we compare our results with other approximate methods for calculating the resistivities and conclude. A description of the numerical details of our implementation and a general outline of the code used can be found in Appendices~\ref{NtotApprox} \&~\ref{code}.

\section{Basic equations}\label{eqs}

The expressions for computing the parallel, perpendicular and Hall resistivities that appear in the generalized Ohm's law \citep{2009ApJ...693.1895K} are
\begin{gather}
\eta_\parallel = \frac{1}{\sigma_\parallel}, \qquad \nonumber \eta_\perp = \frac{\sigma_\perp}{\sigma_\perp^2 + \sigma_H^2}, \\
\eta_H = \frac{\sigma_H}{\sigma_\perp^2 + \sigma_H^2}.
\label{etas}
\end{gather} 
In Eqs.~\ref{etas}, $\sigma_\parallel$, $\sigma_\perp$, and $\sigma_H$ are, respectively, the parallel, perpendicular, and Hall conductivities. The conductivities are, in turn, defined \citep{1991pspa.book.....P} as
\begin{gather}
\sigma_\parallel = \sum_s\sigma_s, \qquad \nonumber \sigma_\perp = \sum_s\frac{\sigma_s}{1+(\omega_s\uptau_{sn})^2}, \\
\sigma_H = -\sum_s\frac{\sigma_s\omega_s\uptau_{sn}}{1+(\omega_s\uptau_{sn})^2},
\label{sigmas}
\end{gather} 
where $\omega_s = e_sB/m_sc$ is the gyrofrequency of species ``$\textit{s}$'', $m_s$ and $e_s$ denote the mass and charge of species $\textit{s}$ respectively, $B$ is the strength of the magnetic field and c is the speed of light. In Eqs.~\ref{sigmas}, $\sigma_s$ is the conductivity of individual species $\textit{s}$ defined as
\begin{equation}\label{sigs}
\sigma_s = \frac{n_se_s^2\uptau_{sn}}{m_s},
\end{equation}
where $n_s$ is the number density of species $\textit{s}$, and $\uptau_{sn}$ is the mean collision time between species $\textit{s}$ and neutral particles, given by
\begin{equation}\label{mct}
\uptau_{sn} = \frac{1}{\alpha_{s\rm{He}}}\frac{m_s+m_{\rm{H_2}}}{\rho_{\rm{H_2}}}\frac{1}{\mean{\sigma w}_{s\rm{H_2}}}.
\end{equation}
In Eq.~\ref{mct}, $m_{\rm{H_2}}$ is the mass of $\rm{H_2}$ particles (the dominant collisional partner of charged species), $\alpha_{sHe}$ is a factor to account for the slowing-down time of species $\textit{s}$ due to the presence and collisions with \rm{He}, $\rho_{\rm{H_2}}$ is the mass density of the cloud, and $\mean{\sigma w}_{s\rm{H_2}}$ is the momentum transfer rate. We assume a constant momentum transfer rate for all ions of $\mean{\sigma w}_{i\rm{H_2}}=2\times{}10^{-9} ~\rm{cm^3~s^{-1}}$ and a constant factor to account for the presence of $\rm{He}$ of $\alpha_{iHe}=1.14$. For electrons we use $\mean{\sigma w}_{e\rm{H_2}}=1.3\times{}10^{-9} ~\rm{cm^3~s^{-1}}$ and $\alpha_{eHe}=1.16$ (see \citeyearless{2008A&A...484...17P}). 

We note here that the mean collisional time between different ions and $\rm{H_2}$ may be different by up to a factor of seven (see Table 1 from \citeyearless{2008A&A...484....1P}). However, assuming a constant momentum transfer rate for all ions is still a very-well justified approximation since, for temperatures relevant to molecular clouds, $\mean{\sigma w}_{i\rm{H_2}}$ can only vary by up to a factor of two between different ions (see Table 2 from \citeyearless{2008A&A...484....1P} and Figs. 3, 5 and 7 from \citeyearless{2008A&A...484...17P}).

In order to calculate the perpendicular, parallel, and Hall resistivities in star-formation simulations, the number densities of charge species, $n_s$, need to be known throughout the modeled molecular cloud, and throughout the evolution of the simulation. In the following sections we make certain simplifying assumptions to reduce the complexity of the above expressions and calculate the conductivities and the resistivities of a cloud, requiring only a few key variables.

\subsection{Parallel conductivity}
From Eqs.~\ref{sigmas} and ~\ref{sigs} the parallel conductivity can be written as
\begin{equation}
    \begin{split}
        \sigma_{\parallel}=\frac{e_s^2}{\rho_{\rm{H_2}}}\sum_s \frac{1}{\alpha_{s\rm{He}}\mean{\sigma w}_{s\rm{H_2}}} \frac{m_s+m_{\rm{H_2}}}{m_s} n_s.
    \end{split}
\end{equation}
Due to their low mass, the contribution of electrons to the parallel conductivity is vastly more important than that of the ions (see Figs. 6-9 in \citeyearless{2022MNRAS.510.4420T}). Therefore, the parallel conductivity can be very well approximated as
\begin{equation}\label{condpar}
    \begin{split}
        \sigma_{\parallel}&\approx\frac{e_s^2}{\rho_{\rm{H_2}}} \frac{1}{\alpha_{e\rm{He}}\mean{\sigma w}_{e\rm{H_2}}} \frac{m_e+m_{\rm{H_2}}}{m_e} n_e \approx5.62\times{}10^{-7} \frac{n_{i,tot}}{\rho_{\rm{H_2}}},
    \end{split}
\end{equation}
where $n_{i,tot}$ is the total number density of all ions which, due to charge neutrality, is equal to the number density of the electrons.

\subsection{Perpendicular conductivity}

To calculate the perpendicular conductivity, we first make the assumption that $\omega_s\uptau_{sn}\gg 1$. This is an excellent approximation for number densities $n_{\rm{H_2}}\le 10^{9} ~\rm{cm^{-3}}$ where both ions and electrons remain very well attached to the magnetic field (see e.g., Fig. 5 from \citeyearless{2022MNRAS.510.4420T} and Fig. 6 from \citeyearless{2007ApJ...660..388T}). With that approximation and using Eq. ~\ref{sigmas}, the perpendicular conductivity can be written as
\begin{equation}
    \begin{split}
        \sigma_{\perp}&\approx \sum_s\frac{\sigma_s}{(\omega_s\uptau_{sn})^2} \approx \frac{c^2\rho_{\rm{H_2}}}{B^2} \sum_s\alpha_{s\rm{He}}\mean{\sigma w}_{s\rm{H_2}}\frac{m_s}{m_s+m_{\rm{H_2}}}n_s.
    \end{split}
\end{equation}
In contrast to the parallel conductivity, the electrons contribute very little compared to the ions (see again Figs. 6-9 of \citeyearless{2022MNRAS.510.4420T}). As such, the perpendicular conductivity can be approximated as
\begin{equation}\label{condperp}
    \begin{split}
        \sigma_{\perp}\approx\frac{1}{C_\perp}\frac{n_{i,tot}}{v_A^2},
    \end{split}
\end{equation}
where $v_A=B/\sqrt{4\pi\rho}$ is the Alfv\'en velocity and $C_\perp$ is a quantity that depends on the chemical properties of a specific cloud. As such, $C_\perp$ essentially encapsulates a weighted average of the contributions of different ions to the perpendicular resistivity, mathematically expressed as 
\begin{equation}\label{WeightedContribution}
    \frac{1}{C_\perp}\approx\frac{c^2}{4\pi}\alpha_{i\rm{He}} \mean{\sigma w}_{i\rm{H_2}} \sum_{s\neq e} \frac{m_s}{m_s+m_{\rm{H_2}}}\frac{n_s}{n_{i,tot}}.
\end{equation}
Since the quantity $m_s/(m_s+m_{\rm{H_2}})$ in the sum does not significantly change between the most dominant ion species (0.86 for $\rm{C^+}$, 0.6 for $\rm{H_3^+}$) we can approximate $C_\perp$ as being dependent only on the physical properties of the cloud and constant throughout the cloud's evolution.

\subsection{Hall conductivity}
Similar to the perpendicular conductivity, to derive a simplified expression for the Hall conductivity, we also assume that $\omega_s\uptau_{sn}\gg1$. With that approximation and taking into account charge neutrality, the expression for the Hall conductivity from Eq.~\ref{sigmas} reads
\begin{equation}
    \begin{split}
        \sigma_H\approx-\sum_s\frac{\sigma_s}{\omega_s\uptau_{sn}} \approx -\frac{c}{B}(e_en_e+e_in_{i,tot})=0.
    \end{split}
\end{equation}
This is to be expected since  $|\sigma_H|\ll|\sigma_{\perp}|$, as also supported by numerical calculations for central number densities below $n_c\le 10^6-10^{7} ~\rm{cm^{-3}}$ (\citeyearless{2002ApJ...573..199N}; \citeyearless{2009ApJ...693.1895K}; \citeyearless{2016A&A...592A..18M}; \citeyearless{2018MNRAS.478.2723Z}). Therefore, for the range of densities our method considers, the Hall conductivity can be ignored with minimal errors to the perpendicular resistivity (assuming a strict case where the Hall conductivity becomes equal to the perpendicular at a density of $10^6~\rm{cm^{-3}}$, the error produced would be ~25\% for the last half order of magnitude where we consider our method to be applicable).

Putting the above approximations in Eqs.~\ref{etas}, the expressions for the resistivities of the cloud can be written as follows:
\begin{gather}
 \eta_{\parallel}=\frac{1}{\sigma_{\parallel}}\approx C_\parallel \frac{\rho_{\rm{H_2}}}{n_{i,tot}}, \qquad \nonumber \eta_{\perp}\approx \frac{1}{\sigma_{\perp}}\approx C_\perp \frac{v_A^2}{n_{i,tot}}, \\
\eta_H\approx 0
\label{etaapproxs}
\end{gather} 
Where $C_\parallel = 1.78 \times 10^6 ~\rm{g^{-1}s}$ and $C_\perp$ depends on the specific physical conditions of a simulated cloud. From Eqs.~\ref{etaapproxs}, it is clear that to calculate the resistivities we only need to know two quantities; the total number density of all ions $n_{i,tot}$ and the constant $C_\perp$. The total ion density is calculated from the total ion abundance $\chi_{i,tot}$, which is approximated as a function of $n_{\rm{H_2}}$ by a power law that evolves alongside the cloud depending on its physical conditions. Both the power law and the constant $C_\perp$ are calculated by interpolations using results from simulations with a detailed chemical network. More details about these calculations are discussed in Appendices~\ref{NtotApprox} and ~\ref{code}.

\begin{figure*}
\centering
\includegraphics[width=\textwidth, clip]{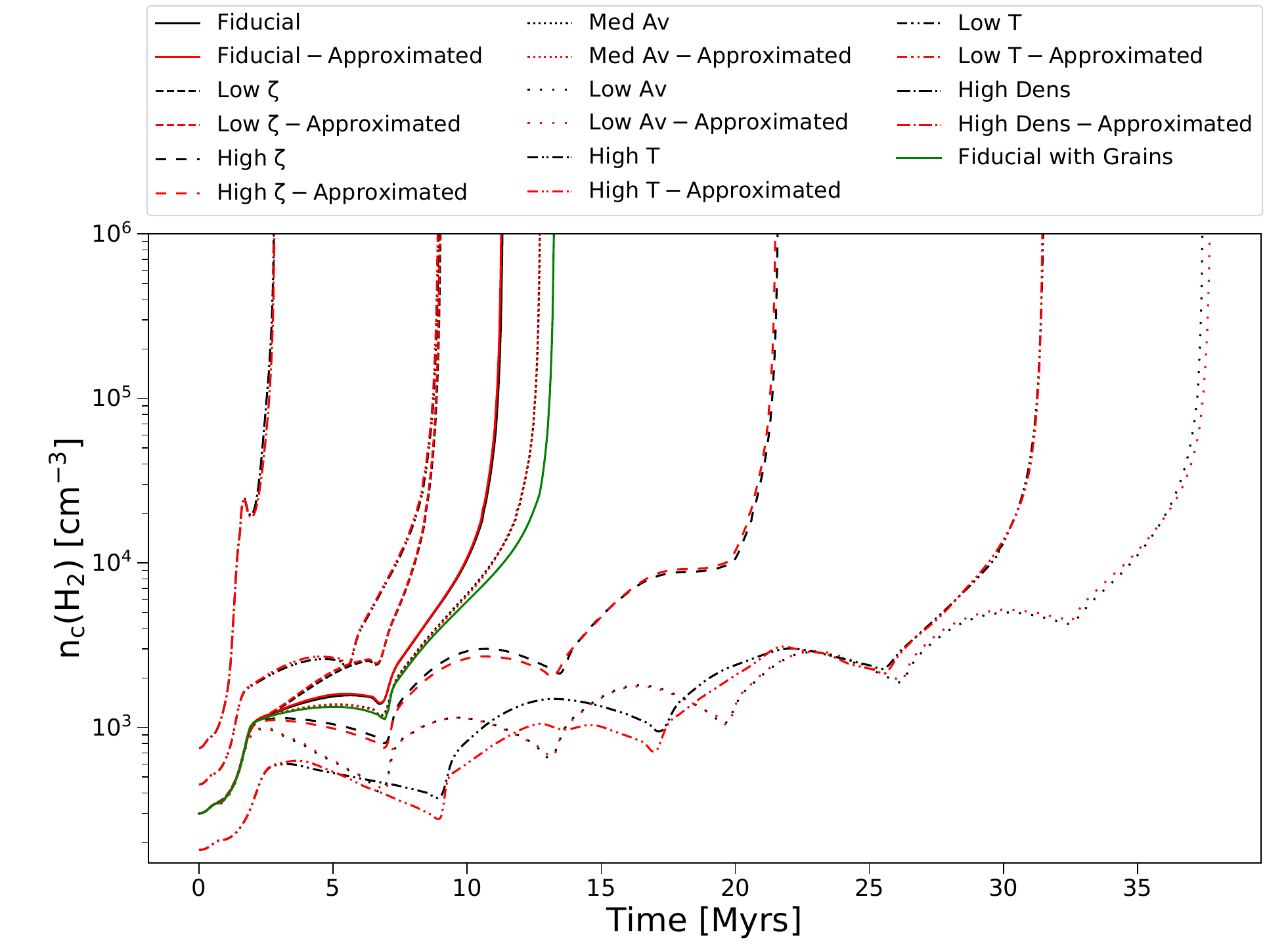}
\caption{\label{cdenssims1} Comparisons of the time evolution of the central $\rm{H_2}$ number density for models used in the interpolation. The black lines represent simulations using a full chemical network and the red lines represent simulations using our approximate method. Different linestyles are used to denote models with different physical parameters (see legend). The green line represents a model with the same physical parameters as the Fiducial model which also includes the effects of grains. Different physical conditions have a more drastic effect in the evolution of the cloud than the inclusion of grains.}
\end{figure*}

\begin{figure*}
\centering
\includegraphics[width=\textwidth, clip]{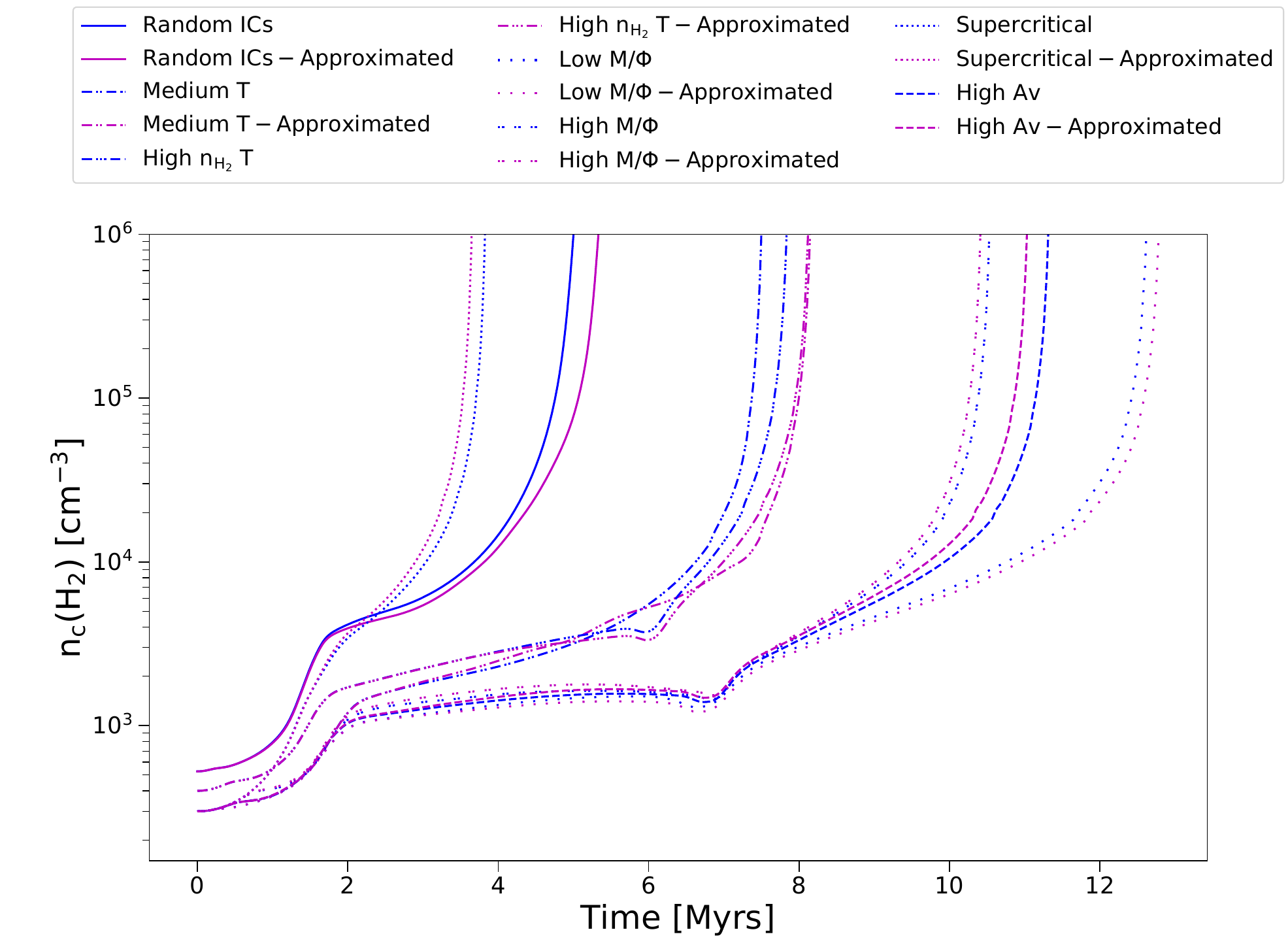}
\caption{\label{cdenssims2} Same as Fig ~\ref{cdenssims1}, but for models where no prior knowledge was present in the interpolating function. The blue lines represent simulations using a full chemical network and the magenta lines represent simulations using our approximate method.}
\end{figure*}

\section{Numerical setup}\label{numsetup}

We follow \cite{2022MNRAS.510.4420T} and \cite{2023MNRAS.521.5087T} to perform a series of 2D non-ideal MHD simulations with non-equilibrium chemistry in cylindrical geometry with radius and half-height of 0.75 pc. Here, we briefly summarize the initial and boundary conditions and refer the reader to \cite{2022MNRAS.510.4420T} for more details. For all models we use an isothermal equation of state, the magnetic field is initially uniform and aligned with the axis of symmetry ($z$-axis), and all velocity components are initially set to zero. For our chemodynamical simulations, we use the chemical network without Deuterium chemistry, first presented in \cite{2022MNRAS.510.4420T}. All of our initial 2D simulations used to develop our interpolation function are performed using an adaptive mesh grid with six levels of refinement and an initial grid size of 32$\times$64 cells in the $r$ and $z$ directions, respectively.

To generalize our method to a large parameter space we simulated various clouds where we altered the cosmic-ray ionization rate, visual extinction, temperature and initial $\rm{H_2}$ density. In Table~\ref{simstable1} we list the physical conditions used in each of these simulated clouds. The mass-to-flux ratio in all these models was equal to 0.5 in units of the critical value for collapse \citep{1976ApJ...210..326M}. In our fiducial model, the cosmic-ray ionization rate is equal to the standard value of $\zeta_0=1.3\times10^{-17} \rm{s^{-1}}$ \citep{1998ApJ...499..234C}, the visual extinction is equal to 10, the temperature is equal to 10 K, and the initial number density is equal to 300 $\rm{cm^{-3}}$.

In order to test the validity of our approximations and the accuracy of the interpolation function developed for an even wider range of physical conditions, we performed additional simulations where we further altered the physical and initial conditions (ICs). However, results from these models never enter the interpolation function. These simulations are listed in the bottom half of Table~\ref{simstable1} and also include models with different initial mass-to-flux ratios. In the interest of computational efficiency, these models are performed considering only one level of refinement, as opposed to the six levels used for the models that enter in the interpolation function. For each model listed in Table~\ref{simstable1}, the resistivities are calculated both by using a non-equilibrium chemical network and via our approximation. Therefore, a total of 30 2D simulations were performed in this study.

Finally, we perform a pair of 3D simulations to test the validity of our method in a 3D simulation as well. A thorough description of these simulations will be provided in a future publication. Consequently, here we focus on the key differences in comparison to the 2D simulations. The cosmic-ray ionization and visual extinction are not constant, but are calculated for every time step and at each grid point inside the computational domain based on a six-ray approximation. Additionally, we include an initial turbulent power spectrum, with $\mathcal{M}=\sigma_v/c_\mathrm{s}=3$ and $\mathcal{M_\mathrm{A}}\approx0.45$. Finally, the value of the mass-to-flux ratio (in units of the critical value for collapse) was set equal to 0.85.

\begin{table*}
\begin{center}
\begin{tabular}{c c c c c c }
\hline\hline
 Interpolation Models & $\zeta$/$\zeta_0$ & $A_v$ [mag] & T [K] & M/$\Upphi$ [(M/$\Upphi)_0$] & $n_{ \rm{H_2},0}\ [\rm{cm^{-3}}]$\\ 
 \hline
 Fiducial  & 1 & 10 & 10 & 0.5 & 300\\  
 Low $\zeta$  & 0.5 & 10 & 10 & 0.5 & 300  \\
 High $\zeta$  & 2 & 10 & 10 & 0.5 & 300  \\
 Medium $A_v$  & 1 & 5 & 10 & 0.5 & 300\\
 Low $A_v$  & 1 & 3 & 10 & 0.5 & 300 \\
 Low T  & 1 & 10 & 6 & 0.5 & 180 \\
 High T  & 1 & 10 & 15 & 0.5 & 450\\
 High $n_{\rm{H_2}}$ & 1 & 10 & 10 & 0.5 & 750\\ 
\hline\hline
Generalized Models & $\zeta$/$\zeta_0$ & $A_v$ [mag] & T [K] & M/$\Upphi$ [(M/$\Upphi)_0$]  & $n_{ \rm{H_2},0}\ [\rm{cm^{-3}}]$\\ 
 \hline
 Random ICs & 1.48 & 8.9 & 11.4 & 0.29 & 526\\
 Medium T &  1 & 10 & 8 & 0.5 & 300\\
 High $n_{\rm{H_2}}$,T  & 1 & 10 & 12.5 & 0.5 & 400\\
 Low M/$\Phi$  & 1 & 10 & 10 & 0.25 & 300\\  
 High M/$\Phi$  & 1 & 10 & 10 & 0.75 & 300  \\
 Supercritical  & 1 & 10 & 10 & 2.6 & 300  \\
 High $A_v$  & 1 & 20 & 10 & 0.5 & 300\\
\hline\hline
\end{tabular}
\end{center}
\caption{\label{simstable1} Top: Parameters of the simulations used in the interpolation function to calculate the resistivities. Bottom: Parameters of the simulations tested blindly. Here, $\zeta_0=1.3\times10^{-17}~\rm{s^{-1}}$ is the standard value of the cosmic-ray ionization rate \citep{1998ApJ...499..234C} and the mass-to-flux ratio is given in units of the critical value \citep{1976ApJ...210..326M}.}
\end{table*}

\begin{figure}
    \centering
    \includegraphics[width=0.49\textwidth, clip]{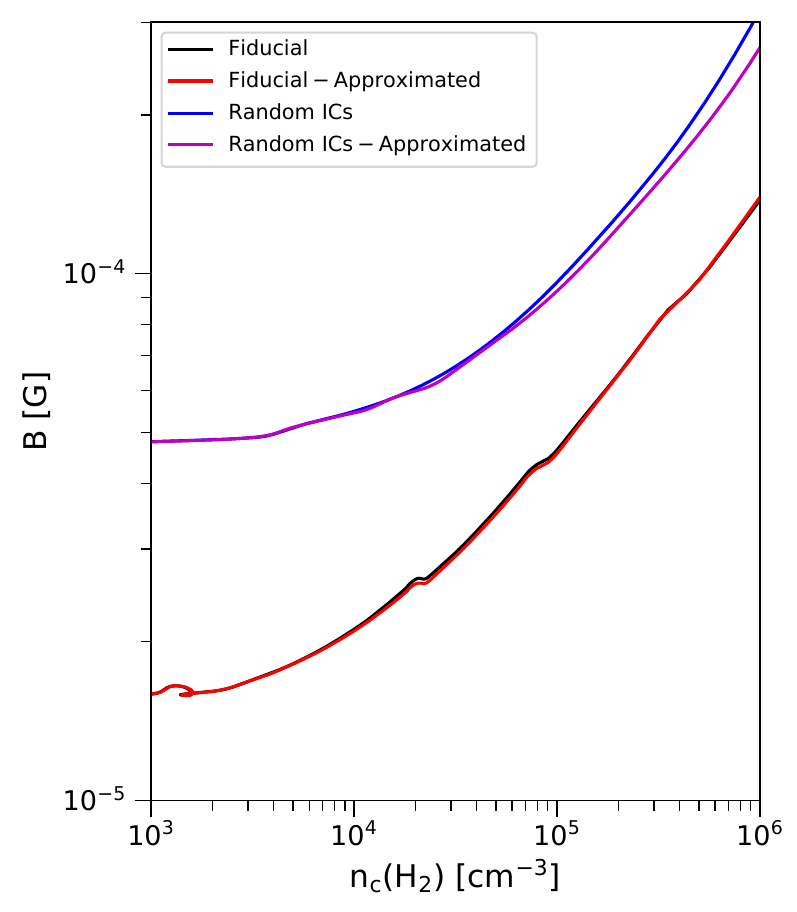}
    \caption{The magnetic field at the center of the cloud as a function of the central $\rm{H_2}$ number density for the fiducial and random ICs models. The black and blue lines represent the fiducial and random ICs simulations respectively using a full chemical network. The red and purple lines represent the fiducial and random ICs simulations respectively using the method described in this paper We can see that the function approaches a power law $B\propto\rho^k$ towards higher densities.}
    \label{B-rho rel}
\end{figure}

\begin{figure*}
\includegraphics[width=1.\textwidth, clip]{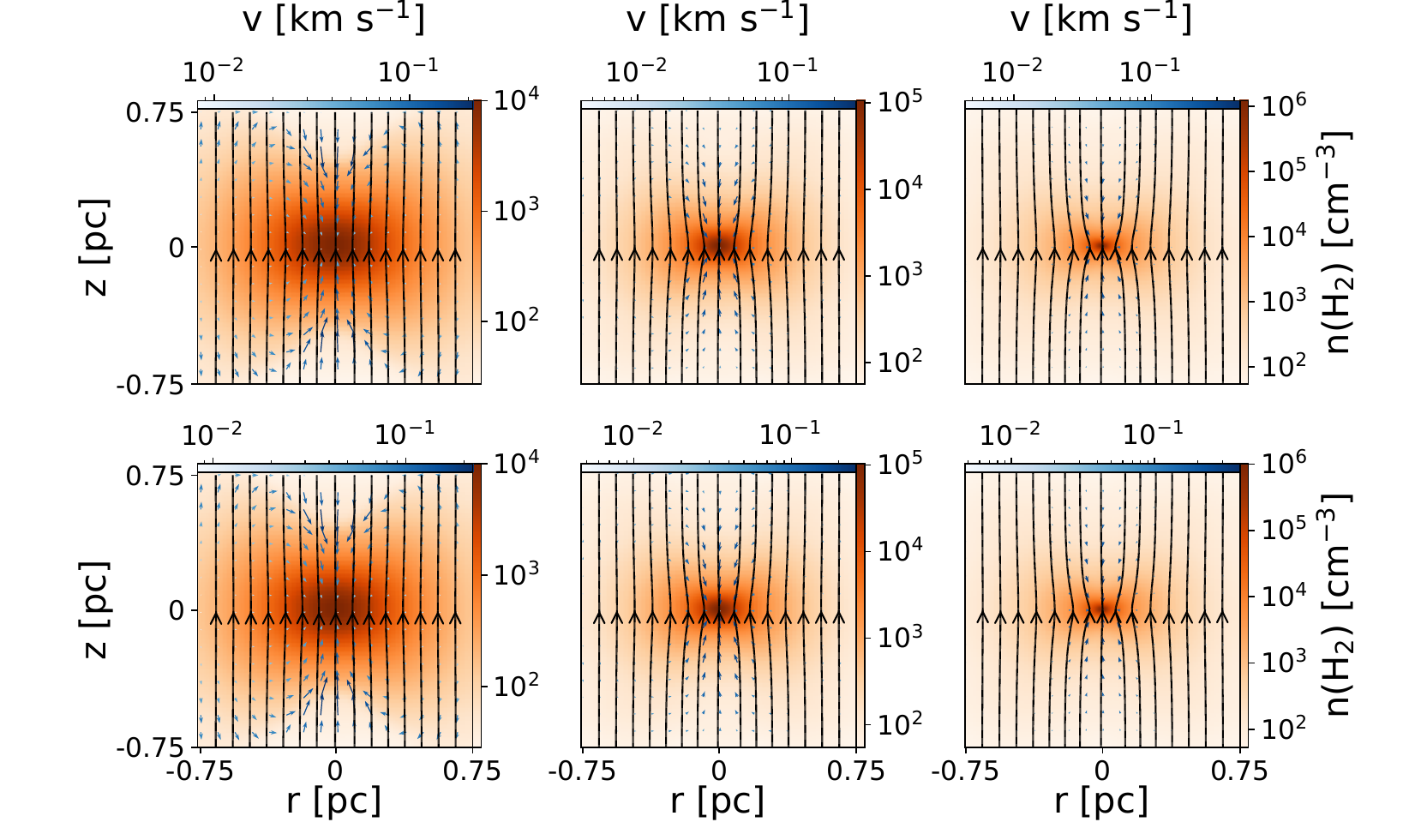}
\caption{Comparison of the spatial distribution of the density between a simulation where the resistivities are calculated using our non-equilibrium chemical network (top row) and a simulation where the resistivities are computed using our approximation (bottom row) for our fiducial model. We compare our results for three different central number densities of $10^4$ (left column), $10^5$ (middle column) and $10^6 ~\rm{cm^{-3}}$ (right column). The black arrows represent the magnetic field and the blue arrows represent the velocity of the plasma.
\label{SpatialComp}}
\end{figure*}

\section{Results}\label{results} 

\subsection{Time evolution-interpolation models}\label{CdensEvol}
In Fig.~\ref{cdenssims1} we show the evolution of the central $\rm{H_2}$ number density of the modeled clouds as a function of time, up to a central number density of $10^6 ~\rm{cm^{-3}}$. Different lines (described in the legend of the figure) show simulations with different physical conditions (see also Table~\ref{simstable1}). With black lines we show our results from the non-ideal MHD simulations with non-equilibrium chemistry, thereby representing the ``ground truth''. With the red lines we plot our results using our newly-developed approximation for calculating the resistivities. As it can be seen from Fig.~\ref{cdenssims1}, the simulations where the resistivities are calculated using our approximation are in excellent agreement with the simulations where the resistivities are calculated from a non-equilibrium chemical network for all physical parameters shown. The agreement, both in regards to the time required for the central $\rm{H_2}$ number density to reach $10^6 ~\rm{cm^{-3}}$, and in terms of the overall evolution of the central density throughout time, shows that our approximation can correctly reproduce non-ideal MHD effects. In other words, the calculation of the resistivities using our approximation leads to an accurate redistribution of magnetic flux throughout the evolution of the cloud.

\subsection{Time evolution-generalized models}\label{CdensBlind}
Fig.~\ref{cdenssims1} shows that our method can reproduce results produced by simulations using non-equilibrium chemistry. However, results from these simulations were already present in our interpolating function. To test the applicability of our method to a large parameter space, in Fig.~\ref{cdenssims2} we show a similar comparison of the evolution of the cloud's central $\rm{H_2}$ density as a function of time for simulations with different parameters tested ``blindly'' (listed in the second half of Table~\ref{simstable1}). That is, no prior knowledge from these chemodynamical was present our interpolating function nor any further optimization was performed in the interpolating function. Here, we have once again changed various physical parameters along with the mass-to-flux ratio. Despite the fact that the mass-to-flux ratio not being considered as a parameter in the interpolation, our approximation still produces very accurate results. Even in the case of the model with random initial conditions, where every parameter is changed simultaneously, our method can still accurately reproduce the results of a simulation with a full chemical network.

\subsection{B $\propto \rho^\kappa$ relation}
Another test to ensure the accuracy of our model is to compare the evolution of the magnetic field at the center of the domain for simulations of various models using non-equilibrium chemistry and the approximate method developed here. In ideal MHD models, the magnetic field evolves with the square root of the density ($B\sim \rho^\kappa$, $\kappa=0.5$; e.g., \citeyearless{2015MNRAS.451.4384T}). In non-ideal MHD models the factor $\kappa$ tends to be slightly lower, depending how efficiently the magnetic flux can be redistributed. As shown in Fig.~\ref{B-rho rel}, the $B-\rho$ relation between simulations using a full chemical network and our model is in very good agreement, for both the fiducial model and the model with random initial conditions. 

\subsection{Spatial comparison}\label{SpatialDistr}

Going beyond the central point of the cloud, for the approximation we have developed to be generally applicable to star formation simulations, it needs to also preserve the morphology and spatial distribution of the density of a simulated cloud as in  
full non-ideal MHD, chemodynamical simulations. To test this, we present in Fig.~\ref{SpatialComp} a comparison between a simulation with full non-equilibrium chemical modeling (top row) and a simulation performed using our approximation (bottom row) for our fiducial model. From left to right columns the central number density of the cloud is $10^4$, $10^5$, and $10^6 ~\rm{cm^{-3}}$, respectively. The number density of the cloud is shown with the orange colormap and the black lines show magnetic field lines, while the velocity field in the cloud is shown with the blue colormap corresponding to the small arrows.

From Fig.~\ref{SpatialComp} it is evident that the simulation where the resistivities are calculated using our approximation is in excellent agreement with the simulation where the resistivities are calculated exactly using the full non-equilibrium chemical network. Specifically, the simulation performed using our approximation yields very accurate results in terms of the morphology and size of the core as well as its kinematic properties.
\begin{figure}
    \centering
    \includegraphics[width=1.\columnwidth, clip]{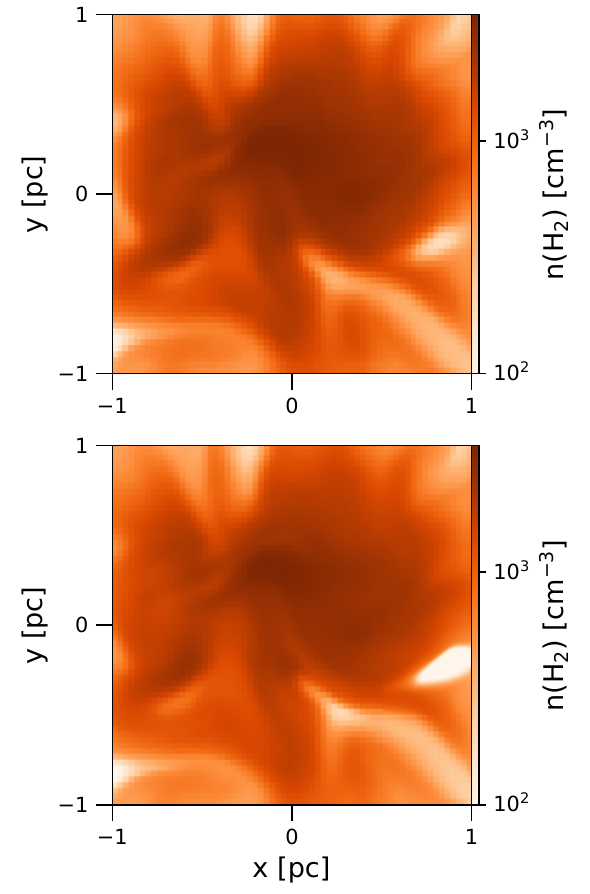}
    \caption{$\rm{H_2}$ density on the central x-y slice for a chemodynamical 3D simulation (top) and a simulation using our method (bottom) at 2.5 Myrs simulation time. The two simulations seem to have evolved very similarly, with only minor differences. The bottom simulation has collapsed slightly slower and the spots where the density is lower are exaggerated compared to the top one.}
    \label{slice3D}
\end{figure}

\subsection{Comparisons of 3-D simulations}\label{3Dsim}

To test how well the approximate method we have developed generalizes to 3D simulations, in Fig. ~\ref{slice3D} we compare the $\rm{H_2}$ number density in the central x-y slice of two 3D simulations. In the upper panel of Fig. ~\ref{slice3D} we present our results from the simulation where the resistivities are calculated using full non-equilibrium chemistry and in the bottom panel we show the same slice using the approximation described herewith. The results from the chemodynamical simulation presented in the upper panel of Fig. ~\ref{slice3D} also include the effect of grains in the calculation of the resistivities, which, we have not taken into account in our approximation. Further discussion about this decision can be found in Sect.~\ref{Grains}.

As evident from Fig. ~\ref{slice3D}, our method approximates the behavior of ambipolar diffusion very well even in a 3D simulation. At a time of 2.5 Myrs, the spatial comparison of the two simulations is in very good agreement. The main differences are that, in the simulation where we used our approximation, the maximum $\rm{H_2}$ density is slightly lower and the underdense region that can be seen in the lower-right part of both images is more prominent. In general however, even with the presence of turbulence and without accounting for the presence of grains, our approximation can accurately reproduce the results from a ``full'' non-ideal MHD chemodynamical simulation, preserving the morphology and fragmentation properties of a cloud.

\section{Discussion}
\subsection{Comparison with other methods}
\begin{figure*}
    \centering
    \includegraphics[width=\textwidth, clip]{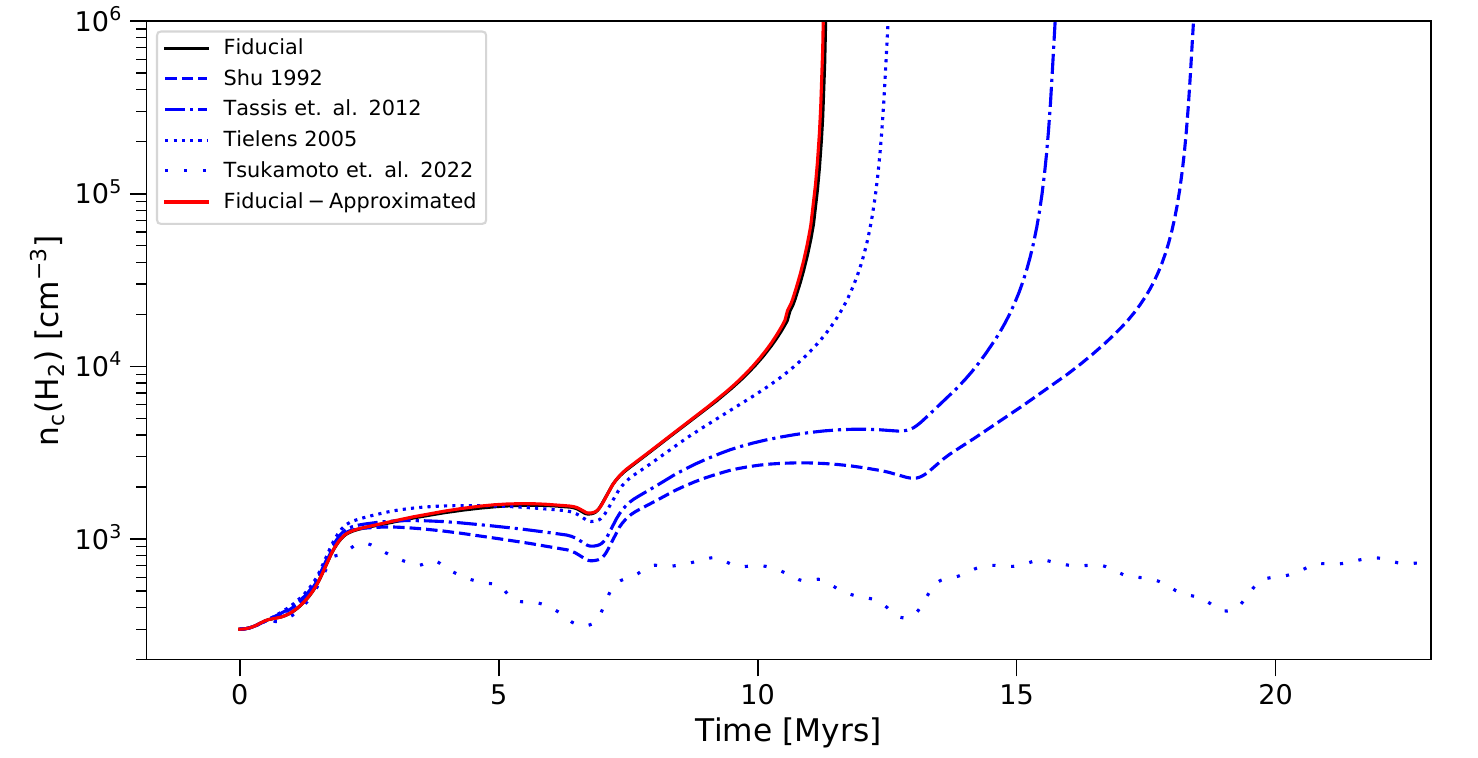}
    \caption{Central $\rm{H_2}$ number density evolution in simulations with identical initial conditions (fiducial model; see Table~\ref{simstable1}) using different methods for calculating the resistivities. The black line represents a simulation using a full chemical network, the red line represents a simulation using the method described in this paper, and the blue lines represent methods proposed in the literature.}
    \label{alts}
\end{figure*}

To demonstrate the accuracy of our implementation in comparison to other approximate methods proposed in the literature, we once again show in Fig.~\ref{alts} the evolution of the central $\rm{H_2}$ number density as a function of time for our fiducial model. The black line shows our results from the full chemodynamical simulation, representing the ``ground truth'', with the red line we show our results from a simulation where the resistivities are calculated using the interpolation function developed here, and with the blue lines we show the results from other approximate expressions for calculating the resistivities.

A common expression used in the literature (e.g., \citeyearless{2024ApJ...961..100A}) for the calculation of the ambipolar-diffusion resistivity is that introduced by \cite{1992pavi.book.....S}. In this approximation, the resistivity is calculated as $\eta_{AD}=\frac{B^2}{4\pi\gamma_{in}\rho_n\rho_i}$, where the density of ions is calculated as $\rho_i=3\times10^{-16}\rho_{\rm{H_2}}^{1/2}$ \citep{1979ApJ...232..729E}. The coupling constant between the ions and the neutrals was taken to be $\gamma_{in} = 3.5\times10^{13} ~\rm{cm^3~g^{-1}~s^{-1}}$. This relation is thought to be valid up to neutral densities of $\sim10^8 ~\rm{cm^{-3}}$ \citep{1987ARA&A..25...23S}, that is in approximately the same density range considered here.

 \cite{2012ApJ...754....6T} (blue dash-dotted line in Fig.~\ref{alts}) used results from models of collapsing cores from \cite{2012ApJ...753...29T} to arrive at an approximation for the ionization fraction of the form $\chi_i=3\times10^{-8}~n_{\rm{H_2}}^{-0.6}$ (see Fig. 5 in \citeyearless{2012ApJ...754....6T}). The models used, simulated cores of radius 0.4 pc, an initial number density of $1000 ~\rm{cm^{-3}}$, temperatures of $\rm{T}=7-15 \rm{K}$, mass-to-flux ratios of $M/\Phi=0.7-1.3~(M/\Phi)_0$, and cosmic-ray ionization rates of $\zeta/\zeta_0=0.25-4$. Therefore, the phase of prestellar-core evolution considered in \cite{2012ApJ...754....6T} (and the resulting expression for the ionization fraction) can be directly compared to the setup considered in our study.
 
 \cite{2005pcim.book.....T} (blue densely dotted line in Fig.~\ref{alts}) considered dense cores ($n_{\rm{H_2}}=10^3-10^8~\rm{cm^{-3}}$) with $A_v\geq$15, temperatures of $\rm{T}\approx10~\rm{K}$ and a cosmic-ray ionization rate of $\zeta/\zeta_0\approx2.3$. The ionization fraction was approximated as $\chi_i\approx\sqrt{\frac{\zeta}{3\times10^{-7}n_{\rm{H_2}}}}$ (see Fig. 10.2 in \citeyearless{2005pcim.book.....T}). Even though the study by  \cite{2005pcim.book.....T} considered somewhat more evolved cores compared to our fiducial case for arriving to an expression for the ionization fraction, their resulting functional form leads to higher values for the ionization fraction and therefore to a longer timescale required for the cloud to collapse.

 \cite{2022arXiv220913765T} (loosely dotted line in Fig.~\ref{alts}) presented three simple power law approximations of $\eta_{AD}$ for three different $\rm{H_2}$ density ranges. For our comparison to their expression, we only used the derived power law corresponding to $\rm{H_2}$ number density lower that $10^7 ~\rm{cm^{-3}}$ ($\eta_{AD}\approx2\times10^{18}\sqrt{\frac{10^{-16}}{\rho_{\rm{H_2}}}}$).

All the approximations we compare our method to are thus aimed at molecular clouds with physical conditions similar to the ones considered in this paper. For the ones that calculate only the ionization fraction, we calculated the number density of ions and inserted it into Eq.~\ref{etaapproxs}. As evident, most methods fall short in terms of reproducing the dynamical evolution of a cloud by several Myrs, while other methods do not even lead to the collapse of the cloud.

\subsection{Effects of grains}\label{Grains}
In Fig. \ref{cdenssims1} we included a model which has the same physical parameters as the fiducial model and includes the effects of grains in the calculation of the resistivities. As is evident from Fig. \ref{cdenssims1}, the effect of grains in the overall time evolution of the cloud is dwarfed in comparison to varying the physical conditions of the cloud.

Previous studies have shown that grains play a significant role in the ionization of a cloud and the resulting resistivities. However, the role of grains most likely becomes dominant for higher densities than the ones considered here. \cite{2023A&A...670A..61M} described the effects of coagulation on the ionization fraction and the resistivities (see their Fig. 7) and demonstrated that these effects become significant for $ \rm{H_2}$ number densities larger than $10^6 ~\rm{cm^{-3}}$. \cite{2018MNRAS.478.2723Z} calculated the resistivites for different grain size distributions. The variation of the ambipolar diffusion resistivity between those different distributions is shown to become significant for $ \rm{H_2}$ number densities larger than $10^5 ~\rm{cm^{-3}}$. The perpendicular conductivity is shown to be larger than the Hall conductivity for densities smaller than $10^6 ~\rm{cm^{-3}}$ regardless of distribution (see their Figs. 6, 7 and 8). 

\cite{2002ApJ...573..199N} found that grains can be the dominant contributor to the perpendicular resistivity for densities larger than $10^4 ~\rm{cm^{-3}}$. However, in their implementation, they assumed a grain size distribution skewed towards smaller grains. Therefore, the vast majority of the grains remained well-attached to the magnetic field at these densities. \cite{2014A&A...572A..20L} and \cite{2015A&A...582A..70S} used observations of the coreshine effect to deduce the grain distribution and found that the maximum grain size in starless cores where a factor of $\sim$5 higher than what is used in \cite{2002ApJ...573..199N} and much closer to values considered in \cite{2009ApJ...693.1895K} and \cite{2016A&A...592A..18M}. The range of densities over which our approximation remains accurate would thus need to be lowered, should any future observations provide evidence for a smaller grain size distribution.

Regardless, such issues, related to the grain size distribution cannot be accurately modeled in any approximate method aiming to model the resistivities. Instead, such complications require a ``full treatment'' of the resistivities where, apart from the ionization fraction, the properties of the grains can also be handled as free parameters. Additionally, given that clouds, either super or sub-critical, spend the majority of their lifetime at lower densities (i.e., $\le10^4 ~\rm{cm{-3}}$), we do not expect that such effects will have a big impact in the overall evolution of the cloud for the range of densities considered here.

Finally, in Sect.~\ref{3Dsim}, we presented a comparison between two 3D simulations, one using detailed chemistry and including the effects of grains and another using our method, which does not take grains into account. The two simulations are in close qualitative and quantitative agreement, with only minor differences in their spatial structure. Due to the reasons described above, we are confident that our method can produce reliable results up to densities of $\sim10^6 ~\rm{cm^{-3}}$, for various physical conditions, even without considering the effect of grains.
\section{Summary}\label{sum}
In this paper we derived approximate expressions for the resistivities that only depend on a handful of variables (magnetic field, $\rm{H_2}$ density, total ion density) and do not require a chemical network to be calculated. Our approximations are very well physically motivated for number densities $n_{\rm{H_2}}\lesssim10^6 ~\rm{cm^{-3}}$. These approximations aim to accurately reproduce the resistivities throughout the entire domain, as opposed to only the ``central'', densest point of a collapsing cloud.

 We carried out a series of simulations using these approximate expressions for the resistivities, which we compare against full non-ideal chemodynamical simulations, simultaneously exploring a large part of the parameter space. We find that our approximation leads to excellent results both in terms of the ambipolar-diffusion timescale, and in terms of the morphology and shape of the cores formed. At the same time, our approximation leads to a factor of $\sim10^2$ increase in computational speed in 2D simulation and $\sim10^4$ for 3D simulations.

In contrast to other approximations aiming at addressing the time step constraint when performing non-ideal MHD simulations (i.e., \citeyearless{2006ApJ...653.1280L}), with the method developed herewith we aim at eliminating the need to include a chemical network. 

We therefore trust that the method developed here for calculating the resistivities can be proven very valuable to the computational star-formation community focusing on the early stages of molecular-cloud evolution and collapsing prestellar cores. The numerical implementation of our method can be found at \href{https://github.com/manosagian/Non-Ideal-MHD-Approximate-Code}{https://github.com/manosagian/Non-Ideal-MHD-Approximate-Code}.

\begin{acknowledgements}
We thank the referee, whose detailed and insightful comments helped improve our manuscript. E. A. and K. T. acknowledge support from the European Research Council (ERC) under the European Unions
Horizon 2020 research and innovation programme under grant agreement No. 7712821. A. Tritsis acknowledges support by the Ambizione grant no. PZ00P2\_202199 of the Swiss National Science Foundation (SNSF). The software used in this work was in part developed by the DOE NNSA-ASC OASCR Flash Center at the University of Chicago. We also acknowledge use of the following software: \textsc{Matplotlib} (Hunter 2007), \textsc{Numpy} (Harris et al. 2020), \textsc{Scipy} (Virtanen et al. 2020), \textsc{Numba} (Lam et al. 2015), and the \textsc{yt} analysis toolkit (Turk et al. 2011). 

\end{acknowledgements}

\appendix
\onecolumn
\section{Approximating the evolution of dominant species}\label{NtotApprox}
\begin{figure*}
    \centering
    \includegraphics[width=0.85\columnwidth, clip]{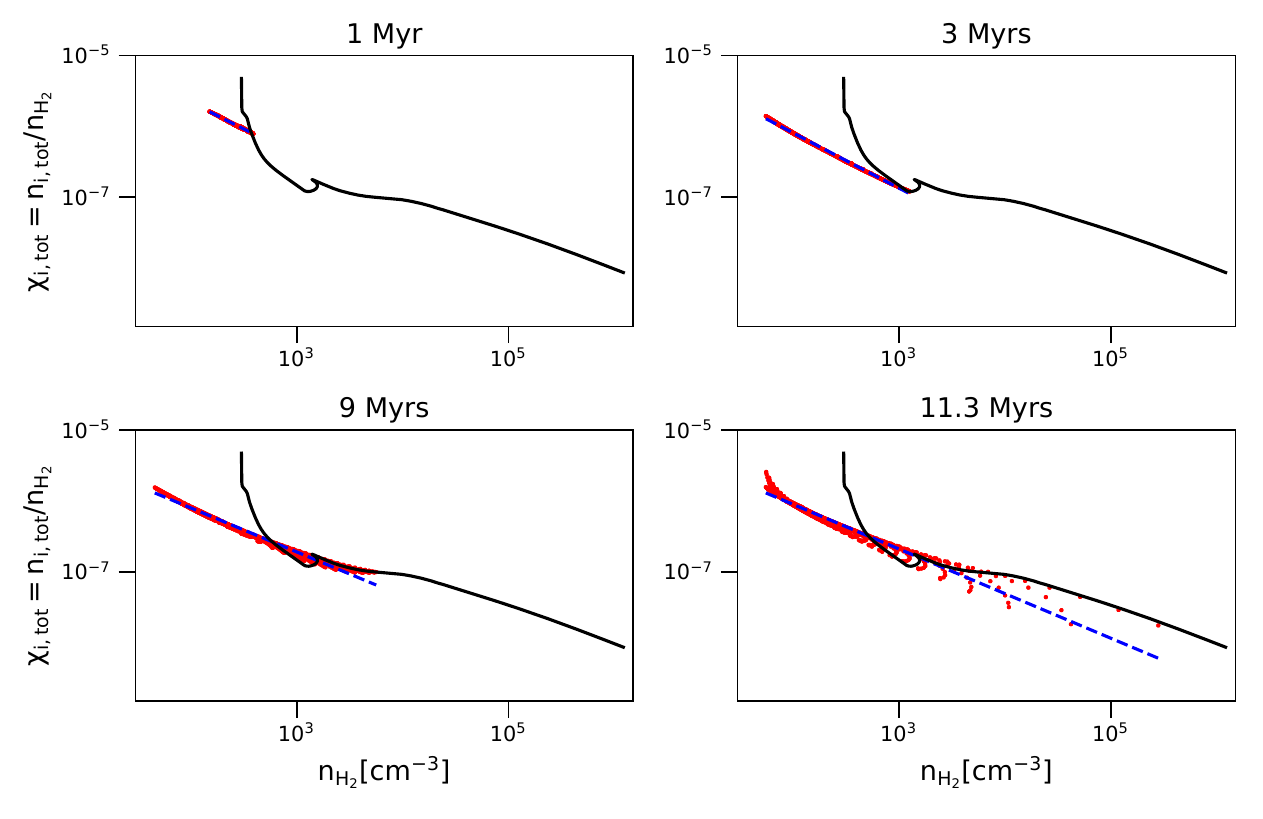}
    \caption{From the top right to bottom left we show the evolution of the ion abundance as a function of the $\rm{H_2}$ number density for the fiducial model at times of 1 Myr, 3 Myrs, 9 Myrs, and 11.3 Myrs. The black line represents the abundance-density relation at the center of the cloud for the entire duration of the simulation, the red points correspond to the ionization fraction from all cells in the domain at the specific point in time, and the blue dashed lines represent the power law that best fits the abundance-density relation at that point in time.}
    \label{ablines}
\end{figure*}
Fig. ~\ref{ablines} shows the relation between the total ion abundance and the $\rm{H_2}$ number density for every point in the domain at different times throughout the evolution of the cloud. Comparing the abundance-$\rm{H_2}$ number density relation for different times, it is evident that the abundance of the dominant species does not follow a single power law with the $\rm{H_2}$ density. Instead, the $\chi_{i,tot}-\rho_{\rm{H_2}}$ relation evolves as the simulation progresses and converges towards a single relation as the cloud collapses. That single relation, however, gives values of $\chi_{i,tot}$ that are up to two orders of magnitude lower than the true values at the beginning of the simulation. An evolving approximation of the abundance is therefore needed.
A simple yet accurate approximation for the abundance, shown in Fig.~\ref{ablines}, is a power law \begin{equation}\label{powerlaw}
    \chi_{i,tot}=A \ (\frac{\rho_{\rm{H_2}}}{\rho_{int}})^B,
\end{equation}
where the constants $\rm{A,B}$ are evolving along with the cloud. The method for approximating the abundance is then the following. We use the geometric mean of the maximum density ($\rho_{max}$) and logarithmic mean of the $\rm{H_2}$ density within the cloud at any time step ($\rho_{int}=\sqrt{\rho_{max,\rm{H_2}} 10^{\mean{\log_{10}{\rho_{\rm{H_2}}}}}}$) to track the evolution of the cloud. This quantity was chosen over simpler ones (e.g., the maximum density of the cloud at any given time), in order to track the cloud's evolution more accurately under turbulent conditions, since our choice of $\rho_{int}$ better reflects the entire distribution of densities over the entire cloud. We then calculate the constants $\rm{A, B}$ as a function of $\rho_{int}$ for the chemodynamical simulations of the models listed in the first half of Table~\ref{simstable1}, which we keep in tabulated form. For a given time step with a given value of $\rho_{int}$, we can then calculate the values of $\rm{A,B}$ at all points in the domain using our multivariate interpolation function that also accounts for the physical conditions (e.g., temperature, visual extinction, etc) of the grid point under consideration. The interpolation function is further explained in Appendix.~\ref{code}.

\section{General code description}\label{code}
The table provided alongside the multivariable interpolation function, consists of 17 columns. The first column (IntDens) consists of the geometric mean of the maximum and logarithmic mean of the $ \rm{H_2}$ density $\rho_{int}$ and the other 8 pairs of columns represent the values of $\rm{A}$ and $\rm{B}$ corresponding to the values of $\rho_{int}$ for the models used in the interpolation. The models used are listed in the first half of Table ~\ref{simstable1} and have varying initial density $n_{\rm{H_2},0}$, temperature T, cosmic-ray ray ionization $\zeta/\zeta_0$ and visual extinction $A_v$. As shown by their names each model focuses on varying one variable compared to the fiducial model. Knowing the values of $\rm{A, B}$ and $\rho_{int}$ , the dominant ion density can then be approximated through a power law as in Eq.~\ref{powerlaw}.

Given the initial $\rm{H_2}$ number density $n_{\rm{H_2},0}$, the value of $\rho_{int}$ at any time step and the values of $\zeta/\zeta_0, A_v, \rm{T}$ of any given point in the computational domain, the constants $\rm{A, B}$ can be interpolated as follows. Firstly, the density $\rho_{int}$ is adjusted so that the interpolation can work for different starting densities. The adjusted density is calculated as
\begin{equation}
    \rho_{adj}=\rho_{int} \frac{n_{Fid}/n_{\rm{H_2},0}+u^2}{1+u^2},
\end{equation}
where u is equal to
\begin{equation}
    u=\frac{log_{10}(\rho_{int}/(2m_pn_{\rm{H_2},0}))}{3n_{\rm{H_2},0}/n_{Fid}}.
\end{equation}

To find the two rows used in the interpolation we need to find the two consecutive rows whose densities bracket the given $\rho_{int}$. To that end, we simply do
\begin{equation}
    \begin{split}
        &for \ i \ in\ [2,size]: \\
        &\ \ if \ IntDens[i]>\rho_{adj}\ then \\
        &\ \ \ \ N=i \\
        &\ \ \ \ stop.
    \end{split}
\end{equation}
The rows used in the interpolation will then be the rows $N$ and $N-1$.

Having selected the rows, we then select the columns (i.e., the models) that will be used in the interpolation. The fiducial model is used as one of the models for the interpolation. For $\zeta/\zeta_0, A_v, \rm{T}$ there are two alternative models where each variable differs from the fiducial. For example, for the cosmic-ray ionization rate $\zeta/\zeta_0$ we have one model with higher and one model with lower rate than the fiducial model's value of $\zeta/\zeta_0=1$. We select which of these two models is more appropriate for the interpolation as follows
\begin{equation}
    \begin{split}
        &If\ (\zeta/\zeta_0 >= 1)\ then \\
 &\ \ f_\zeta = 2 \\
 &\ \ A_{\zeta, N-1} = A_{High\zeta}[N-1] \\
 &\ \ B_{\zeta, N-1} = B_{High\zeta}[N-1] \\
 &\ \ A_{\zeta, N} = A_{High\zeta}[N] \\
 &\ \ B_{\zeta, N} = B_{High\zeta}[N] \\  
 &\ \ C_\zeta = C_{\perp, High\zeta} \\
&Else \\
 &\ \ f_\zeta = 0.5 \\
 &\ \ A_{\zeta, N-1} = A_{Low\zeta}[N-1] \\
 &\ \ B_{\zeta, N-1} = B_{Low\zeta}[N-1] \\
 &\ \ A_{\zeta, N} = A_{Low\zeta}[N] \\
 &\ \ B_{\zeta, N} = B_{Low\zeta}[N] \\  
 &\ \ C_\zeta = C_{\perp, Low\zeta} \\
&End\ If \\
    \end{split}
\end{equation}
where the subscripts denote the model name and the values of the constant $C_\perp$ are given in Table~\ref{consttable}. With this process, we have selected four values $A_{\zeta, N-1}, B_{\zeta, N-1}, A_{\zeta, N}, B_{\zeta, N}$, which correspond to the values of $A, B$ of the appropriate model for the densities bounding $\rho_{int}$. We repeat the selection process for the variables $A_v$ and T. In the models where we varied the temperature we also changed the starting number density. Therefore, the initial number density of the temperature models ($n_T$) also needs to be defined in the selection process and used in the interpolation. Given that there is only one model focused on varying the starting number density of the cloud $n_{\rm{H_2},0}$ (High $n_{\rm{H_2}}$), no selection process is needed for the starting density.

With the interpolation models chosen, we first calculate the constants $\rm{A,B}$ corresponding to the two selected rows, N-1 and N for the given physical conditions. The interpolation function for $\rm{A}$ for the first row (N-1) is of the form
\begin{equation}
\begin{split}
A_{N-1}=&(A_{Fid, N-1} + \frac{n_{Fid}}{n_{\rm{H_2},0}} (A_{\zeta, N-1}-A_{Fid, N-1}) \frac{\zeta-1}{f_\zeta-1}) \\
& (\frac{n_{\rm{H_2},0}}{n_{Fid}})^{\alpha_{A_{N-1}}} 
 (\frac{T}{T_{Fid}})^{\beta_{A_{N-1}}} (\frac{A_{A_v, N-1}}{A_{Fid, N-1}})^{exp_{A_v}},
\end{split}
\end{equation}
where
\begin{equation}
    \begin{split}
&\alpha_{A_{N-1}}=\frac{log_{10}(A_{n_{\rm{H_2},0}, N-1}/A_{Fid, N-1})}{log_{10}(f_{n_{\rm{H_2},0}}/n_{Fid})}\\
&\beta_{A_{N-1}}=\frac{log_{10}(A_{T, N-1}/A_{Fid, N-1})-\alpha_{A_{N-1}} log_{10}(n_T/n_{Fid})}{log_{10}(f_T/T_{Fid})}\\
&exp_{A_v}=\frac{e^{-A_v}-e^{-A_{v_{Fid}}}}{e^{-f_{A_v}}-e^{-A_{v_{Fid}}}}.
    \end{split}
\end{equation}
We repeat that process for the second row (N). To calculate $\rm{A}$, we linearly interpolate between $\rm{A_{N-1}}$ and $\rm{A_N}$
\begin{equation}
    \rm{A}=\rm{A_{N-1}} \frac{\rho_{int}-\rho_{int, N-1}}{\rho_{int,N}-\rho_{int,N-1}}+\rm{A_N} \frac{\rho_{int,N}- \rho_{int}}{\rho_{int,N}-\rho_{int,N-1}}.
\end{equation}
The above process is then repeated for the constant $\rm{B}$.

The constant $C_{\perp}$ used in calculating the perpendicular resistivity is multi-linearly interpolated from the constants in Table~\ref{consttable}, using the same selection process that we used for $\rm{A,B}$. Since $C_{\perp}$ is independent of $\rho_{int}$, the interpolation is performed using only the initial conditions ($\zeta/\zeta_0, A_v, \rm{T}, \rm{and}~ n_{\rm{H_2},0}$) as
\begin{equation}
\begin{split}
    C_{\perp}=&\ C_{Fid}+(C_T-C_{Fid}) \frac{T-T_{Fid}}{f_T-T_{Fid}} \\
    &+(C_{A_v}-C_{Fid}) \frac{A_v-A_{v_{Fid}}}{f_{A_v}-A_{v_{Fid}}}
    +(C_\zeta-C_{Fid}) \frac{\zeta-\zeta_{Fid}}{f_\zeta-\zeta_{Fid}}\\
&+(C_{n_{\rm{H_2},0}}-C_{Fid}) \frac{n_{\rm{H_2},0}-n_{Fid}-
(n_T-n_{Fid}) (T-T_{Fid})/(f_T-T_{Fid})}{f_{n_{\rm{H_2},0}}-n_{Fid}}.
\end{split}
\end{equation}

With the values of $\rm{A}, \rm{B}$, and $C_{\perp}$ calculated, the ion number density can then be computed as in Appendix~\ref{NtotApprox} and the resistivities can be determined as in Sect.~\ref{eqs}.

\begin{table}
\begin{center}
\begin{tabular}{c c}
\hline\hline
 Interpolation Model & $C_{\perp}\ [10^{-12}\ ~\rm{cm^{-5}~s^{3}}]$ \\ 
 \hline
 Fiducial & 7.5\\  
 Low\ $\zeta$ & 7.4\\
 High\ $\zeta$ & 7.0\\
 Medium\ $A_v$ & 7.4\\
 Low\ $A_v$ & 7.3 \\
 Low\ T & 7.8 \\
 High\ T & 7.6 \\
 High\ $n_{\rm{H_2}}$ & 7.2 \\
\hline\hline
\end{tabular}
\end{center}
\caption{\label{consttable} The constant $C_{\perp}$ for the models used in the interpolation}
\end{table}

\end{document}